\documentclass[floatfix,%
preprint,
preprintnumbers,
 amsmath,amssymb,
 aps,
 prab,
]{revtex4-2}
\usepackage{siunitx}
\usepackage{graphicx}
\usepackage{dcolumn}
\usepackage{bm}
\usepackage{xcolor}
\usepackage{amsmath}
\usepackage{booktabs}
\usepackage{hyperref}
\usepackage{tikz}
\usetikzlibrary{shapes.geometric, positioning}
\usepackage{placeins}

\begin{document}


\title{Design criteria for a beam-driven resonant passive transverse deflector for longitudinal beam diagnostics}%

\author{Dmitry Bazyl}
  \email{dmitry.bazyl@desy.de}
  \author{Winfried Decking}
  \author{Sergey Tomin}%
\author{Igor Zagorodnov}%

\affiliation{%
Deutsches Elektronen-Synchrotron DESY, Notkestrasse 85, 22607 Hamburg, Germany
}%

\date{\today}%

\begin{abstract}

Conventional radio-frequency (rf) transverse deflecting structures provide
high-resolution longitudinal beam diagnostics, but require externally generated
high-power rf, waveguide distribution, synchronization and input coupling at
the operating frequency.  We propose design criteria for a beam-driven resonant
passive transverse deflector that does not require an external rf source.  A
leading drive bunch excites long-range wakefields in an off-axis periodic
copper structure and a delayed witness bunch experiences the transverse wake
near a zero crossing.  The concept is based on the large temporal slope available
from high-frequency wake components.
A structure designed for installation after the second bunch compressor in the
three-bunch-compressor layout of the European XFEL is optimized to place the
zero crossing of the drive-bunch-induced transverse wake potential approximately one rf-bucket spacing of the \(\SI{1.3}{GHz}\) linac, behind the drive bunch.  The selected geometry
produces a multi-mode transverse kick dominated by TM-like modes.  We use
time-domain wake simulations, frequency-domain decomposition, cell-number
scaling, mechanical-tolerance scans, orbit-offset studies and uniform thermal
scaling to determine the operating point and its sensitivity.  For this
geometry, the zero crossing occurs at \(s_0\simeq\SI{230.6}{mm}\), with a
per-cell temporal slope of
\(S_{\rm cell}=\SI{1.186}{mV/(pC\,fs\,cell)}\).  For a compact \(\SI{1}{m}\)
structure operated with a \(\SI{250}{pC}\) drive bunch and a \(\SI{700}{MeV}\)
witness beam, the estimated temporal resolution is about \(\SI{33}{fs}\).

\end{abstract}

\maketitle


\section{\label{sec:intr}Introduction}
State-of-the-art high-power radio-frequency (rf) transverse deflecting cavities
are widely used for electron-beam diagnostics
\cite{Panofsky1956TransverseDeflectionRF,Floettmann2014BeamDynamicsTDS,Roehrs2009TimeResolvedTomography,Ding2011FemtosecondXRayTransverseDeflector,Dolgashev2014MultimegawattXBandDeflectors,Behrens2014FewFemtosecondXFEL,Craievich2020PolariX,Marchetti2021PolariXBeamCharacterization,GonzalezCaminal2024PolariXCommissioning,Prat2025AttosecondPolariXSwissFEL}.
Such cavities are operated in externally driven dipole modes and require
high-power rf infrastructure and stable synchronization.  In contrast, the
method proposed in Ref.~\cite{Tomin2026BeamDrivenTDS} relies on beam-induced
resonant electromagnetic fields, or wakefields and omits the need of high-power rf.  In this concept, a leading drive bunch excites resonant
fields in a periodic iris-loaded copper rf structure. A delayed witness
bunch samples the corresponding transverse kick.  In general, the drive bunch excites
a spectrum of resonant modes, with high-frequency components suppressed by the
bunch form factor.  Through geometric optimization, the modal contributions can
be phased such that they add constructively at the witness-bunch delay.  The effective streaking voltage increases approximately linearly with
the number of periods in the deflecting structure over the range considered
here.

The proposed beam-driven transverse deflecting structure is closely
related to earlier work on passive wakefield streaking and plasma-wakefield
diagnostics.  Passive corrugated or dechirper-based structures \cite{Bane2012} have used
self-induced short-range wakefields to correlate longitudinal position with
transverse kick
\cite{Bettoni2016WakefieldStreaking,Seok2018CorrugatedPassive,Dijkstal2022SelfSynchronized, dijkstal2024longitudinal}.
In plasma-based diagnostics, laser-driven plasma wakes have been proposed as
attosecond streaking fields, while drive--probe experiments have measured
transverse plasma wakefields excited by an off-axis leading bunch and sampled
by a trailing witness bunch
\cite{Dornmair2016PlasmaDiagnostic,Lindstrom2018TransverseWakefields}.
Related high-frequency deflector work by Dolgashev and collaborators includes
W-band and millimeter-wave structures in which beam-driven fields and
deflecting gradients were studied
\cite{Dolgashev2015WbandDeflectors,DalForno2016MMWaveDeflecting}.
The distinguishing feature of the proposed resonant approach is the use of sufficiently
long-lived resonant wakefields excited by a preceding drive bunch, allowing a
delayed witness bunch to sample an approximately linear zero-crossing kick.  In
this sense, the structure acts as a passive analogue of an rf transverse
deflector.

The following sections develop the modal design criterion and apply it to a finite
off-axis copper structure.  We first define the zero-crossing operating point,
the spatial and temporal wake slopes and the modal overlap quantities.  The
closed-pillbox model is then used to estimate the offset and aperture
dependence of the dominant TM-like modes.  \mbox{Time-domain} wake simulations
are used to determine the optimized geometry, the per-cell slope, the
cell-number scaling and the harmonic content of the wake derivative.
Mechanical, orbit and temperature sensitivity studies quantify the stability of the operating point, while a low-energy European XFEL example illustrates its practical implementation for beam diagnostics.

\section{Modal principle and design criteria}
\label{sec:modal-principle}

\begin{figure}[!htb]
  \centering
  \includegraphics[width=0.6\linewidth]{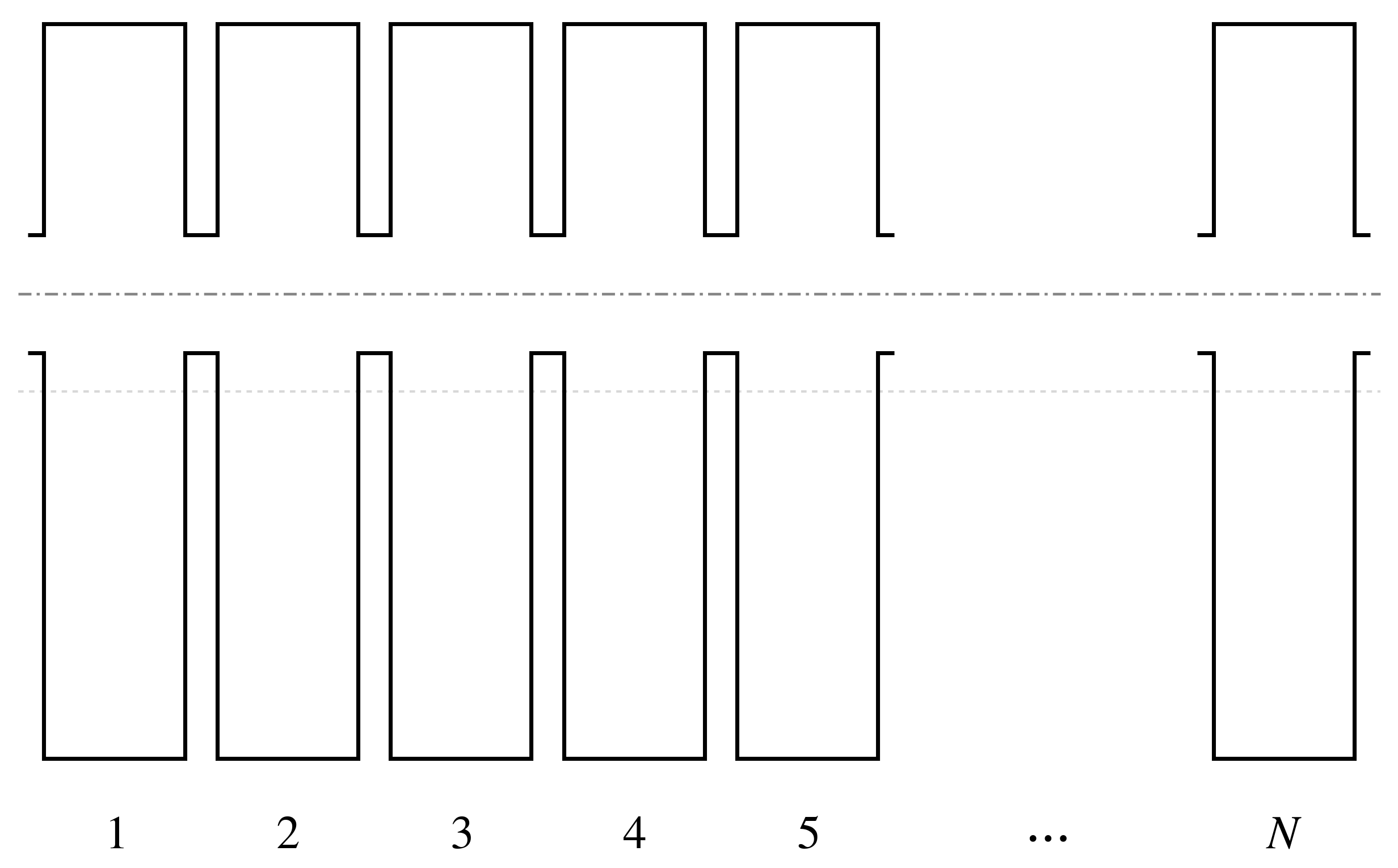}
  \caption{
  Schematic longitudinal section of an \(N\)-cell off-axis pillbox array.
  The light dashed line denotes the cylindrical pillbox axis.  The darker
  dash-dotted line denotes the displaced beam trajectory through the beam
  pipe.  The relevant geometric parameters are the pillbox radius \(R\), cell
  length \(L_c\), period \(P\), beam-pipe diameter \(d_{\rm pipe}\) and
  transverse offset \(\rho=|\boldsymbol{\rho}|\).
  }
  \label{fig:ncell-schematic}
\end{figure}

Consider a drive bunch of charge \(q_d\) travelling with velocity
\(v\simeq c\) through a periodic \(N\)-cell passive deflecting structure,
shown schematically in Fig.~\ref{fig:ncell-schematic}.  The structure is
characterized by the cell period \(P\), pillbox radius \(R\), cell length
\(L_c\), beam-pipe diameter \(d_{\rm pipe}\) and transverse pipe-offset
vector \(\boldsymbol{\rho}\) measured from the pillbox axis to the beam-pipe
axis.  A witness bunch follows the drive bunch at a longitudinal distance
\(s>0\).  The wake definitions are written for a general cross-wake
configuration, in which the drive and witness bunches may have different
transverse positions \(\mathbf r_d\) and \(\mathbf r_w\).  The same-trajectory
case used for the design formulas is obtained by setting
\(\mathbf r_d=\mathbf r_w=\mathbf r_b\); for the nominal design trajectory,
\(\mathbf r_b=\boldsymbol{\rho}\).

The coordinate system is right-handed, with \(z\) along the nominal bunch
trajectory, \(x\) in the horizontal plane and \(y\) in the vertical plane.
The distance \(s\) denotes the drive--witness separation: a field
left behind by the drive bunch is sampled by the witness at time
\(t=(z+s)/c\).  The wake potentials are defined using the standard convention
for relativistic beams~\cite{Chao1993CollectiveInstabilities,ZotterKheifets1998,Stupakov2018WakeImpedance}.
The longitudinal wake potential per unit drive charge is defined as
\begin{equation}
  W_{\parallel}(\mathbf r_w,\mathbf r_d;s)
  =
  -\frac{1}{q_d}
  \int_{\Gamma_w}
  E_z\!\left(
    \mathbf r_w,z,t=\frac{z+s}{c}
  \right)
  dz ,
  \qquad s>0 .
  \label{eq:longitudinal-wake}
\end{equation}
Here \(\Gamma_w\) is the witness trajectory.  With this convention, a positive
longitudinal wake corresponds to energy loss of a trailing electron bunch when
\(q_d\) is understood as the positive charge magnitude of the drive bunch.

The transverse wake potential is defined from the transverse Lorentz voltage
per unit drive charge,
\begin{equation}
  \mathbf W_{\perp}(\mathbf r_w,\mathbf r_d;s)
  =
  \frac{1}{q_d}
  \int_{\Gamma_w}
  \left[
    \mathbf E_{\perp}
    +
    c\,\hat{\mathbf z}\times\mathbf B
  \right]_
  {\mathbf r_w,z,t=(z+s)/c}
  dz ,
  \qquad s>0 .
  \label{eq:transverse-wake}
\end{equation}
The sign of the actual transverse momentum change is obtained by multiplying
this voltage by the witness charge.  The selected streaking direction is
denoted by the transverse unit vector \(\mathbf e_s\) and the projected
transverse wake is
\begin{equation}
  W_s(\mathbf r_w,\mathbf r_d;s)
  =
  \mathbf e_s\cdot
  \mathbf W_{\perp}(\mathbf r_w,\mathbf r_d;s).
  \label{eq:projected-transverse-wake}
\end{equation}
For vertical streaking \(\mathbf e_s=\hat{\mathbf y}\), whereas for horizontal
streaking \(\mathbf e_s=\hat{\mathbf x}\). Later, the structure optimization, scaling and sensitivity studies use the vertical wake. For this case \(W_y\equiv W_s\) and \(D_y\equiv D_s\) and the corresponding temporal slope is denoted by \(S_y\).

The passive deflector is operated near a zero crossing of the long-range
transverse wake,
\begin{equation}
  W_s(\mathbf r_w,\mathbf r_d;s_0)\simeq 0,
  \qquad
  D_s(\mathbf r_w,\mathbf r_d;s_0)
  =
  \left.
  \frac{dW_s(\mathbf r_w,\mathbf r_d;s)}{ds}
  \right|_{s=s_0}.
  \label{eq:zero-crossing-derivative}
\end{equation}
If the longitudinal coordinate inside the witness bunch is denoted by
\(\zeta\), measured as an additional delay relative to the reference separation
\(s_0\), then
\begin{equation}
  W_s(\mathbf r_w,\mathbf r_d;s_0+\zeta)
  =
  W_s(\mathbf r_w,\mathbf r_d;s_0)
  +
  D_s(\mathbf r_w,\mathbf r_d;s_0)\,\zeta
  +
  O(\zeta^2).
  \label{eq:local-linear-wake}
\end{equation}
At an exact zero crossing, \(W_s(\mathbf r_w,\mathbf r_d;s_0)=0\) and the
local wake derivative \(D_s\) is the passive analogue of the deflecting-voltage
slope in an active transverse deflecting rf structure.  In this paper \(D_s\)
denotes the spatial derivative with respect to the drive--witness separation
\(s\).  If the corresponding time coordinate is defined by \(s=ct\), the
temporal streaking slope is
\begin{equation}
  S_y(\mathbf r_w,\mathbf r_d;s_0)
  =
  \left.
  \frac{dW_y(\mathbf r_w,\mathbf r_d;s)}{dt}
  \right|_{s=s_0}
  =
  cD_y(\mathbf r_w,\mathbf r_d;s_0).
  \label{eq:time-spatial-derivative}
\end{equation}

For a relativistic drive and witness bunch, the long-range wake is represented
below as a sum over resonant eigenmodes, following the standard wakefield and
impedance formalism~\cite{Chao1993CollectiveInstabilities,ZotterKheifets1998}.
Such an eigenmode expansion is exact for a closed, lossless perfectly
conducting cavity, where the modes form a complete orthogonal set and have
real eigenfrequencies.  The present structure, however, is open and may include
material or external losses.  In this case the representation by real
resonance frequencies and quality factors is an approximation to the more
general pole expansion of the impedance.

In the more accurate resonator treatment, the wake is obtained from the complex
poles of the impedance.  The damping is then accompanied by a shift of the
oscillation frequency, as discussed in the standard treatment of resonator
wakes and impedances by Chao~\cite{Chao1993CollectiveInstabilities}.  In the
present work we neglect this small frequency shift and use the simpler
engineering representation in which each relevant resonance is described by a
real resonance frequency \(\omega_m\) and a loaded quality factor \(Q_m\).
This approximation is sufficient because the purpose of this section is
to formulate the modal design principle and the dominant mode-selection
criteria.

We use the field convention \(\exp(-i\omega t)\).  For mode \(m\), let
\(\omega_m\) be the real angular resonance frequency used in this approximate
modal representation, \(Q_m\) the corresponding loaded quality factor and
\begin{equation}
  k_m=\frac{\omega_m}{c},
  \qquad
  \alpha_m=\frac{\omega_m}{2Q_m c}
  \label{eq:modal-k-alpha}
\end{equation}
the synchronous wavenumber and spatial damping constant, respectively.

From this point on we specialize the working modal formulas to the
same-trajectory case,
\(\mathbf r_d=\mathbf r_w=\mathbf r_b\) and
\(\Gamma_d=\Gamma_w=\Gamma_b\).  The synchronous longitudinal voltage of mode
\(m\) on this common trajectory is
\begin{equation}
  V_{z,m}(\mathbf r_b)
  =
  \int_{\Gamma_b}
  E_{z,m}(\mathbf r_b,z)
  \exp(ik_m z)\,dz ,
  \label{eq:drive-longitudinal-voltage}
\end{equation}
The Gaussian drive-bunch form factor is
\begin{equation}
  F_m
  =
  \exp\!\left[
    -\frac{1}{2}
    \left(
      \frac{\omega_m\sigma_z}{c}
    \right)^2
  \right],
  \label{eq:gaussian-form-factor}
\end{equation}
where \(\sigma_z\) is the rms bunch length.

With \(U_m\) denoting the stored energy of mode \(m\), the positive
longitudinal modal amplitude is defined by the standard same-trajectory
cavity-mode loss-factor normalization,
\begin{equation}
  K_{\parallel,m}(\mathbf r_b)
  =
  \frac{
    \left|V_{z,m}(\mathbf r_b)\right|^2
  }{2U_m}
  F_m .
  \label{eq:positive-longitudinal-amplitude}
\end{equation}
This expression is independent of the arbitrary normalization of the eigenmode
fields: the voltages scale linearly with the field amplitude, while the stored
energy scales quadratically.  In the common higher-order mode convention,
\(K_{\parallel,m}=2k_m^{\rm loss}F_m\).  The longitudinal modal wake is then
written as
\begin{equation}
  W_{\parallel,m}(\mathbf r_b;s)
  =
  K_{\parallel,m}(\mathbf r_b)
  \exp(-\alpha_m s)
  \cos(k_m s),
  \qquad s>0 .
  \label{eq:longitudinal-modal-wake}
\end{equation}

The transverse Lorentz voltage of mode \(m\), projected onto the selected
streaking direction, is defined by
\begin{equation}
  V_{s,m}(\mathbf r_b)
  =
  \mathbf e_s\cdot
  \int_{\Gamma_b}
  \left[
    \mathbf E_{\perp,m}
    +
    c\,\hat{\mathbf z}\times\mathbf B_m
  \right](\mathbf r_b,z)
  \exp(ik_m z)\,dz .
  \label{eq:witness-transverse-voltage}
\end{equation}
The corresponding positive transverse modal amplitude is
\begin{equation}
  K_{s,m}(\mathbf r_b)
  =
  \frac{
    \left|
      V_{z,m}^*(\mathbf r_b)
      V_{s,m}(\mathbf r_b)
    \right|
  }{2U_m}
  F_m .
  \label{eq:positive-transverse-amplitude}
\end{equation}
The transverse modal wake in the selected streaking direction
is then
\begin{equation}
  W_{s,m}(\mathbf r_b;s)
  =
  K_{s,m}(\mathbf r_b)
  \exp(-\alpha_m s)
  \sin(k_m s),
  \qquad s>0 .
  \label{eq:transverse-modal-wake}
\end{equation}
Thus the longitudinal wake of a resonant mode has the cosine phase, while the
transverse wake has the sine phase.

The phase relation follows from the Panofsky--Wenzel theorem
\cite{Panofsky1956TransverseDeflectionRF,Chao1993CollectiveInstabilities,Stupakov2018WakeImpedance}.
With the wake definitions used above,
\begin{equation}
  \frac{\partial
  \mathbf W_\perp(\mathbf r_w,\mathbf r_d;s)}
  {\partial s}
  =
  \nabla_{\perp,w} W_\parallel(\mathbf r_w,\mathbf r_d;s) ,
  \label{eq:panofsky-wenzel}
\end{equation}
where the transverse gradient is taken with respect to the witness position,
with the drive position held fixed.  In the same-trajectory specialization,
\(\mathbf r_w=\mathbf r_d=\mathbf r_b\) is imposed only after this gradient is
evaluated.
For the same harmonic convention, after this order of operations is applied,
the corresponding voltage relation can be written, apart from end-point terms,
as
\begin{equation}
  \mathbf V_{\perp,m}(\mathbf r_b)
  =
  \frac{i}{k_m}
  \nabla_{\perp} V_{z,m}(\mathbf r_b).
  \label{eq:pw-voltage-form}
\end{equation}
Here \(\nabla_{\perp} V_{z,m}(\mathbf r_b)\) is shorthand for the
witness-coordinate gradient specified in Eq.~\eqref{eq:panofsky-wenzel},
evaluated on the common trajectory.  Thus the transverse Lorentz voltage is
determined by the transverse gradient of the synchronous longitudinal voltage
on that trajectory.

Summing over modes gives
\begin{equation}
  W_{\parallel}(\mathbf r_b;s)
  =
  \sum_m
  K_{\parallel,m}(\mathbf r_b)
  \exp(-\alpha_m s)
  \cos(k_m s),
  \qquad s>0 ,
  \label{eq:longitudinal-modal-sum}
\end{equation}
and
\begin{equation}
  W_s(\mathbf r_b;s)
  =
  \sum_m
  K_{s,m}(\mathbf r_b)
  \exp(-\alpha_m s)
  \sin(k_m s),
  \qquad s>0 .
  \label{eq:transverse-modal-sum}
\end{equation}

The exact local derivative of Eq.~\eqref{eq:transverse-modal-sum} is
\begin{equation}
  D_s(\mathbf r_b;s_0)
  =
  \sum_m
  K_{s,m}(\mathbf r_b)
  \exp(-\alpha_m s_0)
  \left[
    k_m\cos(k_m s_0)
    -
    \alpha_m\sin(k_m s_0)
  \right].
  \label{eq:modal-derivative-exact}
\end{equation}
For weak damping over the drive--witness delay,
\(\alpha_m s_0\ll 1\), this reduces to
\begin{equation}
  D_s(\mathbf r_b;s_0)
  \simeq
  \sum_m
  k_m K_{s,m}(\mathbf r_b)
  \cos(k_m s_0).
  \label{eq:modal-derivative-weak-damping}
\end{equation}
The condition \(\alpha_m s_0\ll 1\) is equivalent to
\begin{equation}
  Q_m \gg \frac{\omega_m s_0}{2c}.
  \label{eq:weak-damping-condition}
\end{equation}

Equations~\eqref{eq:positive-longitudinal-amplitude}
and~\eqref{eq:positive-transverse-amplitude} give the central design
criterion.  A mode contributes efficiently to the passive deflection only if
it has strong longitudinal coupling to the common beam trajectory and strong
transverse Lorentz voltage, or equivalently a strong transverse gradient of
the synchronous longitudinal voltage on that trajectory.  This is in contrast
to an active rf transverse deflecting structure, where the external rf system
directly excites the design deflecting mode.  In the beam-driven case, an
off-axis bunch can couple efficiently to modes whose closed-cavity parents are
not dipole deflecting modes.  TM-like modes
contribute significantly when their longitudinal voltage is large on the
common trajectory and varies strongly in the selected streaking direction. The useful streaking slope is determined by the coherent modal sum in
Eq.~\eqref{eq:modal-derivative-exact}, or by
Eq.~\eqref{eq:modal-derivative-weak-damping} in the weak-damping limit.  

\subsection{Mode selectivity}
\label{sec:parent-mode-scaling}

For a closed pillbox of radius \(R\) and length \(L_c\), the
TM parent frequencies are
\begin{equation}
  f_{mnq}
  =
  \frac{c}{2\pi}
  \sqrt{
    \left(\frac{\chi_{mn}}{R}\right)^2
    +
    \left(\frac{q\pi}{L_c}\right)^2
  },
  \label{eq:pillbox-frequency}
\end{equation}
where \(\chi_{mn}\) is the \(n\)-th zero of \(J_m\) and \(q\) is the axial
index.  

For a mode with
\begin{equation}
  E_{z,mnq}
  =
  E_0
  J_m\!\left(\chi_{mn}\frac{r}{R}\right)
  \mathcal C_{m\mu}(\theta)
  Z_q(z),
  \label{eq:parent-ez}
\end{equation}
where \(\mu\) labels the two angular polarizations for \(m>0\), Maxwell's
equations give
\begin{equation}
  \mathbf H_{\perp,mnq}
  \propto
  \hat{\mathbf z}\times\nabla_\perp E_{z,mnq}.
  \label{eq:parent-h-from-ez}
\end{equation}
Consequently,
\[
  \hat{\mathbf z}\times\mathbf B_{\perp,mnq}
  \propto
  -\nabla_\perp E_{z,mnq},
\]
up to a mode-dependent normalization and phase.  The magnetic part of the
transverse Lorentz force is governed by the transverse gradient of
the same \(E_z\) field that excites the mode.

Let
\begin{equation}
  \Phi_{mn\mu}(r,\theta)
  =
  J_m\!\left(\chi_{mn}\frac{r}{R}\right)
  \mathcal C_{m\mu}(\theta)
  \label{eq:parent-phi}
\end{equation}
denote the transverse dependence of the parent longitudinal field.  The
magnetic contribution to the drive--witness overlap scales as
\begin{equation}
  \mathcal G^{(B)}_{s,mn\mu}
  (\mathbf r_d,\mathbf r_w)
  =
  \frac{2-\delta_{m0}}
       {R^2J_{m+1}^2(\chi_{mn})}
  \,
  \Phi_{mn\mu}(\mathbf r_d)
  \left[
    -\mathbf e_s\cdot
    \nabla_\perp \Phi_{mn\mu}(\mathbf r_w)
  \right].
  \label{eq:magnetic-parent-overlap-general}
\end{equation}
The first factor is the longitudinal drive coupling, the second is the
magnetic Lorentz kick projected onto the streaking direction.  In the
collinear geometry, \(\mathbf r_d=\mathbf r_w=\boldsymbol{\rho}\).

For radial scaling estimates one may suppress the angular factor and keep only
the dependence on \(u=\chi_{mn}\rho/R\).  The overlap then contains the Bessel
product
\begin{equation}
  \mathcal G^{(B)}_{mn}
  \propto
  (2-\delta_{m0})
  \frac{
    \chi_{mn}J_m(u)J'_m(u)
  }{
    R^3J_{m+1}^2(\chi_{mn})
  },
  \qquad
  u=\chi_{mn}\frac{\rho}{R}.
  \label{eq:magnetic-radial-overlap}
\end{equation}
The corresponding parent-mode
estimate for the wake amplitude is
\begin{equation}
  A_{mnq}
  \propto
  \mathcal G^{(B)}_{mn}
  |T_q(\omega_{mnq})|^2
  F_{mnq},
  \label{eq:parent-amplitude-scaling}
\end{equation}
where \(T_q\) is the longitudinal transit factor and \(F_{mnq}\) is the drive
bunch form factor. 

The first maxima of \(|J_m(u)J'_m(u)|\) for the lowest radial families are
listed in Table~\ref{tab:offset-optima}.  The offset ratio can therefore bias
the coupling toward different TM modes.

\begin{table}[!htb]
  \centering
  \caption{
  First maxima of the radial parent-mode product
  \(|J_m(u)J'_m(u)|\), with \(u=\chi_{m1}\rho/R\).
  }
  \label{tab:offset-optima}
  \begin{tabular}{c c c}
    \toprule
    Angular order \(m\) & \(u_{\rm max}\) & \((\rho/R)_{\rm max}\) \\
    \midrule
    0 & 1.08 & 0.450 \\
    1 & 0.91 & 0.238 \\
    2 & 2.06 & 0.402 \\
    3 & 3.15 & 0.493 \\
    \bottomrule
  \end{tabular}
\end{table}

For the monopole parent modes,
\begin{equation}
  J_0(u)J'_0(u)
  =
  -J_0(u)J_1(u),
  \label{eq:monopole-product}
\end{equation}
and the first maximum occurs near
\begin{equation}
  u\simeq1.08,
  \qquad
  \left(\frac{\rho}{R}\right)_{\rm max}
  \simeq
  \frac{1.08}{\chi_{01}}
  \simeq
  0.45 .
  \label{eq:monopole-offset-optimum}
\end{equation}
For example, an
off-axis TM\(_{010}\)-like mode, commonly used as the accelerating mode, has large longitudinal drive voltage and a
nonzero radial magnetic field at the witness trajectory.  It can therefore
contribute to a transverse streaking wake.

Near the cavity axis,
\begin{equation}
  J_0(u)J'_0(u)\simeq -\frac{u}{2},
  \qquad
  J_m(u)J'_m(u)\sim u^{2m-1}
  \quad
  (m\ge 1).
  \label{eq:small-offset-scaling}
\end{equation}
The monopole and \(m=1\) TM modes therefore turn on linearly with
offset, whereas higher angular orders are suppressed at small \(\rho/R\).

\subsection{Beam-pipe aperture and modal bandwidth}
\label{sec:pipe-cutoff-design}

The beam-pipe aperture is a primary design constraint.  For a circular pipe of
radius \(a=d_{\rm pipe}/2\), the lowest dipole-like cutoff is
\begin{equation}
  f_{c,\mathrm{TE}_{11}}
  =
  \frac{1.84118\,c}{2\pi a},
  \label{eq:te11-cutoff}
\end{equation}
and the first monopole TM cutoff is
\begin{equation}
  f_{c,\mathrm{TM}_{01}}
  =
  \frac{2.40483\,c}{2\pi a}.
  \label{eq:tm01-cutoff}
\end{equation}
A smaller beam pipe raises these cutoff frequencies and can confine a larger
set of high-frequency TM-like components.  Such components can increase the
available streaking slope through the factor \(k_m\) in the local derivative,
provided that they are not strongly suppressed by the drive-bunch form factor
and that their phases remain useful at the witness delay. In contrast, accelerator
operation favors a larger aperture for beam transport, alignment
tolerance and reduced sensitivity to orbit offsets.


\section{Optimized multi-cell structure and modal evidence}
\label{sec:larger-five-cell-geometry}

\begin{figure}[!htb]
  \centering
  \includegraphics[width=0.65\linewidth]{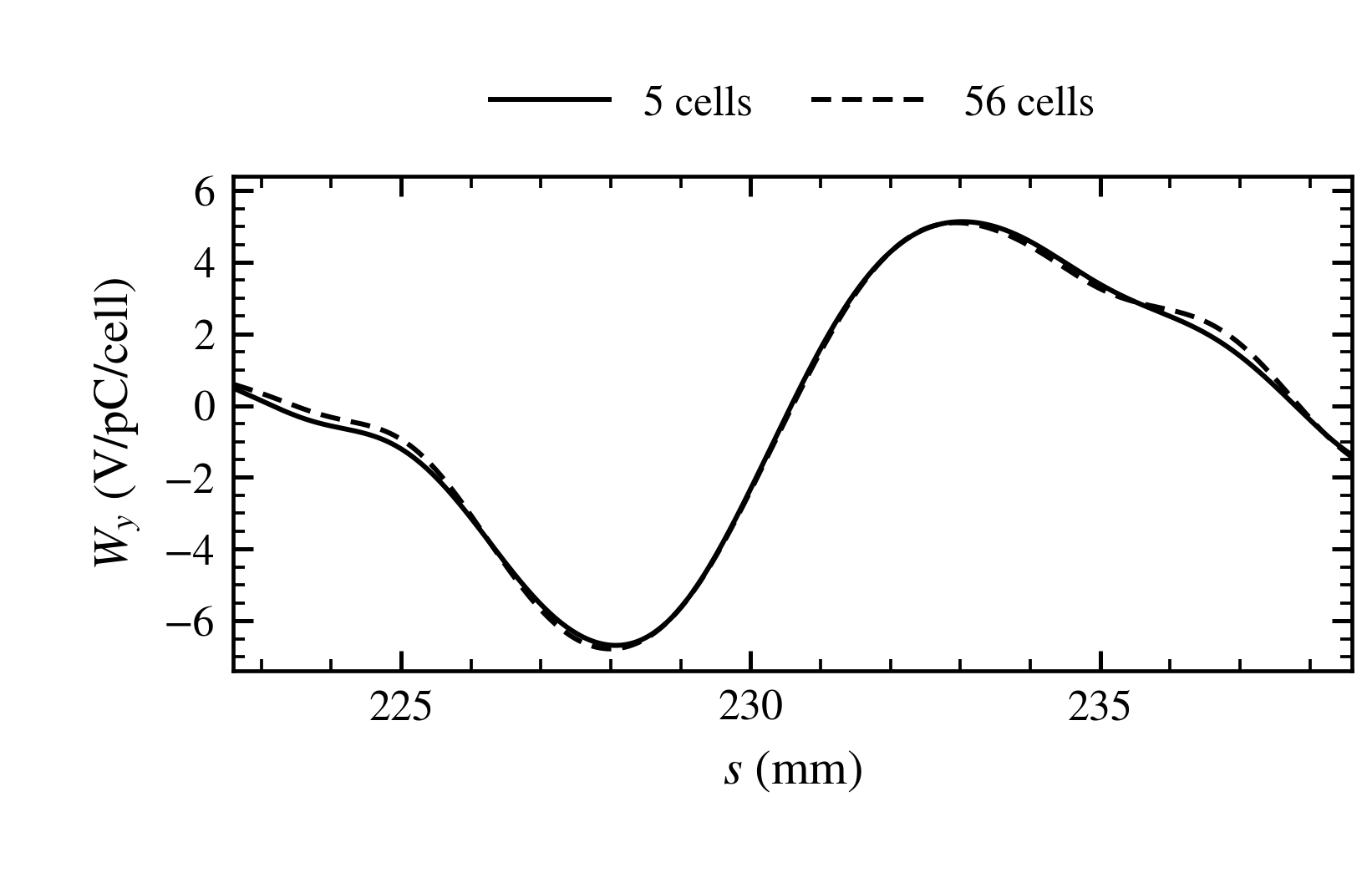}
  \caption{
  Transverse wake potential per cell for the 5-cell and 56-cell structures,
  shown near the selected zero crossing \(s_0=\SI{230.6}{mm}\).  The wake
  potentials are divided by the corresponding number of cells before plotting,
  giving the ordinate in \(\si{V/(pC\,cell)}\).  The close agreement of the two
  traces indicates approximately linear scaling of the transverse wake with
  the number of cells over this simulated range.
  }
  \label{fig:wake-scaling-per-cell}
\end{figure}

The geometry analyzed below was obtained by genetic algorithm optimization of the
off-axis pillbox parameters.  The objective function was the magnitude of the
spatial transverse-wake derivative at one \(\SI{1.3}{GHz}\) rf-bucket
separation,
\begin{equation}
  \mathcal F_{\rm opt}
  =
  \left|
  \frac{dW_y}{ds}
  \right|_{s=s_{\rm rf}},
  \qquad
  s_{\rm rf}
  =
  \frac{c}{1.3~\mathrm{GHz}}
  =
  \SI{230.6}{mm}.
  \label{eq:optimization-objective}
\end{equation}
The electromagnetic simulations were performed with CST Studio
Suite~\cite{CSTStudioSuite}.  The time-domain wakefield calculations use the
finite-integration technique~\cite{Weiland1977FIT} and were used for the optimization, cell-number scaling, tolerance scans,
and orbit-offset studies.  The harmonic amplitudes and phases used in the decomposition were obtained from the time-domain wake potential.  The mode classification is assigned by comparison with closed-pillbox analytic frequency estimates and eigenmode field
patterns for the same optimized geometry.
The optimized geometric parameters are listed in
Table~\ref{tab:optimized-geometry}.  The resulting wake has a zero crossing
close to the target delay.  To compare structures of different length, we use
the temporal slope per cell, \(S_{\rm cell}=S_y/N\), evaluated at the selected
zero crossing.  Figure~\ref{fig:wake-scaling-per-cell} compares the 5-cell and
56-cell wakes after dividing each wake by its number of cells.  The close
agreement supports approximately linear scaling of the temporal slope with cell
number over the simulated range.  The small residual difference between the
per-cell wakes may reflect finite-length effects and remaining numerical
uncertainty in the wake calculation.  A mesh convergence study showed a weak
numerical-dispersion shift at moderate mesh resolution, which was reduced by
increasing the mesh density.

\FloatBarrier
\begin{table}[t]
  \centering
  \caption{
  Geometric parameters of the optimized 5-cell off-axis pillbox structure.
  }
  \label{tab:optimized-geometry}
  \begin{tabular}{l c c}
    \toprule
    Parameter & Symbol & Value \\
    \midrule
    Pillbox radius & \(R\) & \(\SI{9.38}{mm}\) \\
    Cell length & \(L_c\) & \(\SI{3.59}{mm}\) \\
    Period & \(P\) & \(\SI{4.42}{mm}\) \\
    Beam-pipe diameter & \(d_{\rm pipe}\) & \(\SI{3.01}{mm}\) \\
    Pipe offset & \(\rho\) & \(\SI{2.48}{mm}\) \\
    Active-cell fill factor & \(L_c/P\) & \(0.81\) \\
    Offset ratio & \(\rho/R\) & \(0.265\) \\
    \bottomrule
  \end{tabular}
\end{table}

For \(d_{\rm pipe}=\SI{3.01}{mm}\), Eqs.~\eqref{eq:te11-cutoff} and
\eqref{eq:tm01-cutoff} give
\(f_{c,\mathrm{TE}_{11}}=\SI{58.4}{GHz}\) and
\(f_{c,\mathrm{TM}_{01}}=\SI{76.3}{GHz}\).  The closed-pillbox
frequency estimates for pillbox radius \(R=\SI{9.38}{mm}\) are
\(f_{010}\simeq\SI{12.2}{GHz}\),
\(f_{110}\simeq\SI{19.5}{GHz}\) and
\(f_{210}\simeq\SI{26.1}{GHz}\).  The
\(\SI{3}{mm}\)-class beam pipe was chosen for reliable beam transport and
alignment margin in the low-energy European XFEL section.  The drive bunch has
an rms length
\(\sigma_z=\SI{80}{\micro m}\).  For this bunch length the Gaussian
form factor is close to unity over the resolved modal spectrum: it is
\(F=0.9841\) at \(\SI{106.68}{GHz}\).  The mode damping due to lossy copper (\(\sigma_{\rm Cu}=\SI{5.8e7}{S/m}\)) is
included in the time-domain wake calculation and in the eigenmode analysis and
optimization.  For the dominant eigenmodes of the optimized cell, the
perturbative wall-loss quality factors are
\(Q_m\simeq4.4\times10^3\)--\(6.7 \times10^3\) over
\(\SI{12}{GHz}\le f_m\le\SI{28}{GHz}\).  At
\(s_0=\SI{230.596}{mm}\), this corresponds to
\(\exp(-\alpha_m s_0)=0.993\)--\(0.989\), so wall-loss damping changes these
modal amplitudes by only \(0.7\%\)--\(1.1\%\) over the witness delay, or one
linac rf-bucket spacing.

The optimized offset ratio, \(\rho/R=0.265\), lies closest to the first
maximum of the \(m=1\) TM parent mode in Table~\ref{tab:offset-optima}. It
is well below the monopole maximum near \(\rho/R\simeq0.45\) and below the
\(m=2\) maximum near \(\rho/R\simeq0.40\).  Compared with the geometry of
Ref.~\cite{Tomin2026BeamDrivenTDS}, which had \(\rho/R\simeq0.4\), the
parent-mode scaling therefore predicts a weaker TM\(_{010}\)-like contribution
and a stronger role for TM-like modes dominated by the \(m=1\) parent family.
In the finite off-axis structure the displaced beam pipe breaks cylindrical
symmetry, so the simulated modes are generally mixtures of several closed-pillbox
parent families rather than pure \(m\) modes.

The optimized wake has a useful zero crossing near one rf-bucket spacing,
\begin{equation}
  s_0\simeq\SI{230.6}{mm}.
  \label{eq:large-geometry-wake-zero}
\end{equation}
A local linear fit to the simulated 5-cell wake \(W_y(s)\) around this zero
crossing, converted according to \(S_y=c\,dW_y/ds\), gives
\[
  S_y(s_0)
  =
  \SI{5.931}{mV/(pC\,fs)} .
\]
Since the wake scaling with cell number was verified separately over the
5-cell and 56-cell simulations, we quote the streaking strength per cell,
\begin{equation}
  S_{\mathrm{cell}}(s_0)
  =
  \frac{1}{N}
  S_y(s_0)
  =
  \SI{1.186}{mV/(pC\,fs\,cell)},
  \qquad N=5 .
  \label{eq:large-geometry-wake-result}
\end{equation}

For a structure with \(N\) coherently contributing cells, let \(u_s\) denote the
screen coordinate in the selected streaking plane.  The corresponding  streaking factor is
\begin{equation}
  \frac{du_s}{dt}
  =
  R_s\,
  \frac{q_d}{E_w}\,
  N S_{\rm cell},
  \label{eq:streaking-calibration}
\end{equation}
where \(q_d\) is the drive-bunch charge expressed in the same charge unit used
for the wake normalization, \(E_w\) is the witness-beam energy in
electron-volts and \(R_s\) is the transfer matrix element from the structure
to the screen in the streaking plane (\(R_s=R_{12}\) and
\(R_s=R_{34}\) for horizontal and vertical streaking respectively).  For the quoted \(S_{\rm cell}\), \(q_d\)
is therefore given in pC.  With \(E_w\) expressed in electron-volts and the
wake voltage in volts, the transverse voltage divided by \(E_w\) gives the
dimensionless kick angle.  The corresponding rms temporal resolution is

\begin{equation}
  \sigma_{t,\rm res}
  =
  \frac{
    E_w\,\sigma_{u_s,0}
  }{
    |R_s|\,q_d\,N\,|S_{\rm cell}|
  },
  \label{eq:temporal-resolution}
\end{equation}
with \(\sigma_{u_s,0}\) the unstreaked rms beam size on the screen in the
diagnostic plane.

\begin{figure}[!htb]
  \centering
  \includegraphics[width=0.65\linewidth]{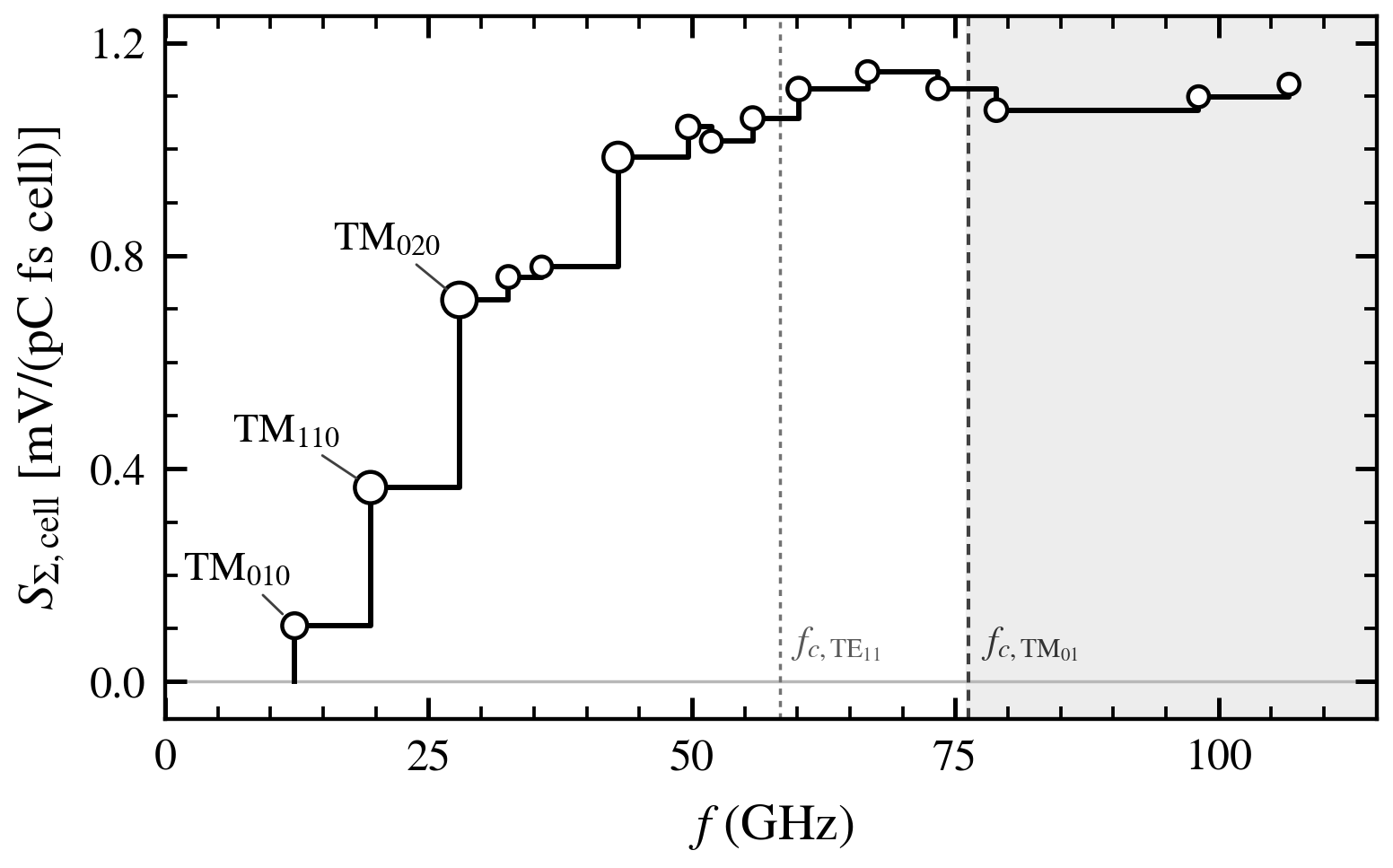}
  \caption{
  Wake-only harmonic decomposition of the local temporal transverse-wake slope
  for the nominal 5-cell structure.  The staircase shows the cumulative
  resolved per-cell contribution
  \(S_{\Sigma,\mathrm{cell}}(f)=\sum_{f_i\le f}S_{i,\mathrm{cell}}\) at the
  witness zero crossing.
  Marker area is proportional to the absolute individual contribution
  \(|S_{i,\mathrm{cell}}|\).  The vertical lines mark the TE$_{11}$ and
  TM$_{01}$ circular beam pipe cutoffs.  The shaded region denotes frequencies
  above the TM$_{01}$ cutoff.  The mode labels indicate the closest ideal
  closed-pillbox TM parent families.
  }
  \label{fig:nominal-wake-decomposition}
\end{figure}

The resolved harmonic content of the optimized 5-cell
wake is shown in Fig.~\ref{fig:nominal-wake-decomposition} and detailed in Table~\ref{tab:wake-slope-contributions}. The harmonic content is consistent with a periodic eigenmode
interpretation.  Floquet eigenmodes of a single period can be used to
reconstruct the finite-structure response by weighting the modes with the
appropriate phase advance and finite-cell coherence factor.  For the geometry
considered in Ref.~\cite{Tomin2026BeamDrivenTDS}, this reconstruction was
evaluated from finite multi-cell eigenmode data and showed good agreement with
the direct time-domain wake calculation.  For optimization, the time-domain
wake calculation is the most direct approach because the resulting wake
potential already contains the coherent sum of the relevant modal
contributions.

\begin{table}[t]
  \centering
  \caption{
  Wake-derived modal contributions to the local temporal streaking slope at
  the zero crossing.  The decomposition is normalized per cell.  Parent-mode
  labels denote ideal closed-pillbox modes.
  }
  \label{tab:wake-slope-contributions}
  \begin{tabular}{cccc}
    \toprule
    \(f_i\) [GHz] &
    parent family &
    \(S_{i,\rm cell}\) [mV/(pC\,fs\,cell)] &
    \(S_{i,\rm cell}/S_{\rm cell}\) \\
    \midrule
    27.946 & TM$_{020}$-like & +0.323 & 27.2\% \\
    19.499 & TM$_{110}$-like & +0.246 & 20.8\% \\
    42.999 & TM$_{220}$-like & +0.215 & 18.1\% \\
    12.291 & TM$_{010}$-like & +0.106 &  8.9\% \\
    26.027 & TM$_{210}$-like & +0.101 &  8.5\% \\
    49.649 & TM$_{320}$-like & +0.060 &  5.1\% \\
    60.178 & TM$_{040}$-like & +0.040 &  3.4\% \\
    55.751 & TM$_{420}$-like & +0.042 &  3.6\% \\
    78.914 & TM$_{331}$-like & -0.037 & -3.1\% \\
    73.286 & TM$_{430}$-like & -0.024 & -2.1\% \\
    \midrule
    --- & other resolved components & +0.087 &  7.3\% \\
    --- & unresolved residual & +0.028 &  2.4\% \\
    \midrule
    --- & total & +1.186 & 100.0\% \\
    \bottomrule
  \end{tabular}
\end{table}

\section{\label{sec:optimization}Sensitivity analysis}

\begin{figure}[!htb]
  \centering
  \includegraphics[width=\linewidth]{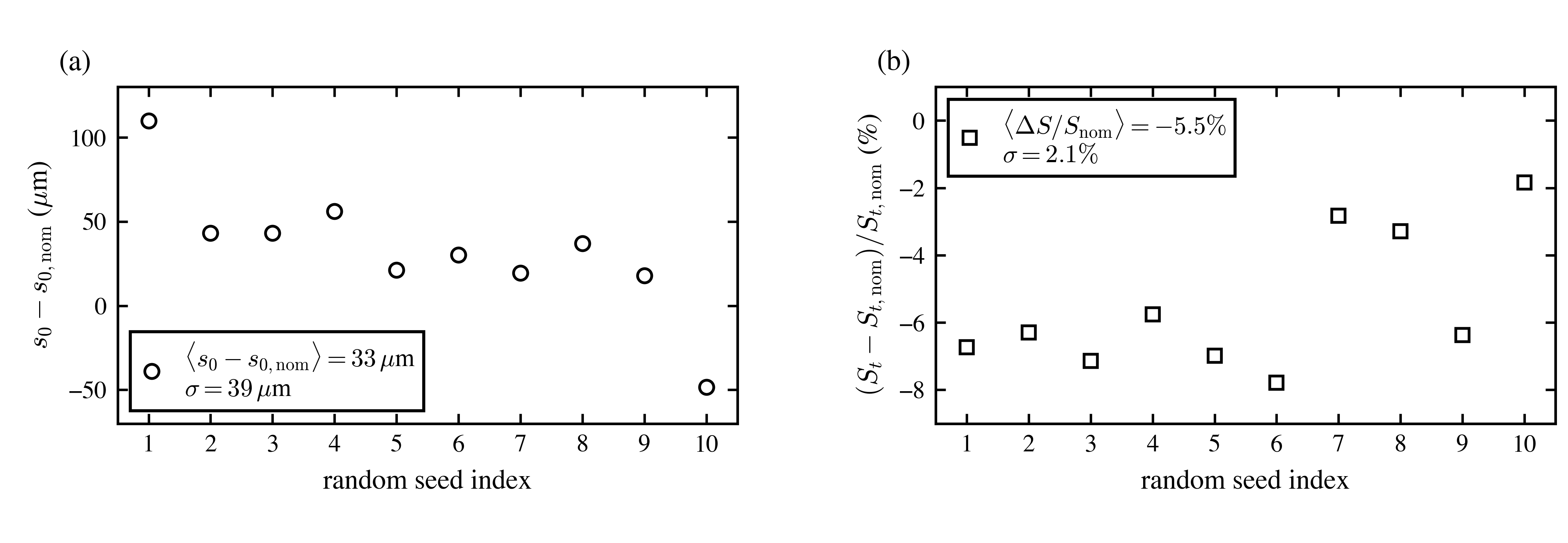}
  \caption{
  Mechanical-error sensitivity of the transverse wake zero crossing.  Ten
  independently seeded structures were generated with uncorrelated Gaussian
  cell-to-cell perturbations of rms amplitude \(\SI{10}{\micro\meter}\).  (a)
  Shift of the nearest transverse-wake zero crossing after translating the
  longitudinal coordinate so that the nominal error-free structure is aligned
  to the design witness delay.  (b)~Relative change of the local streaking
  slope evaluated at each case's own zero crossing.
  }
  \label{fig:mechanical-error-tolerance}
\end{figure}

Mechanical errors were modeled for a \(\SI{250}{mm}\)-long structure
(56 cells) by applying uncorrelated Gaussian cell-to-cell perturbations with
moderate rms amplitude of \(\SI{10}{\micro m}\).  Ten random seeds were
simulated.  For each seed, the nearest zero crossing to the nominal witness
delay was found and the local slope was evaluated at that crossing.
Figure~\ref{fig:mechanical-error-tolerance} shows a sample-mean slope
reduction of \(\SI{5.5}{\percent}\) with a standard deviation of
\(\SI{2.1}{\percent}\).  The zero-crossing shift has a sample mean of
\(+\SI{33}{\micro m}\) and a standard deviation of \(\SI{39}{\micro m}\), with
all tested cases remaining within about \(\SI{110}{\micro m}\) of the nominal
crossing.  This level of phase shift requires beam-based calibration of the
operating phase. However, the tested cell-to-cell perturbations do not eliminate the operating zero crossing or the local
slope. 

\begin{figure}[!htb]
  \centering
  \includegraphics[width=0.65\linewidth]{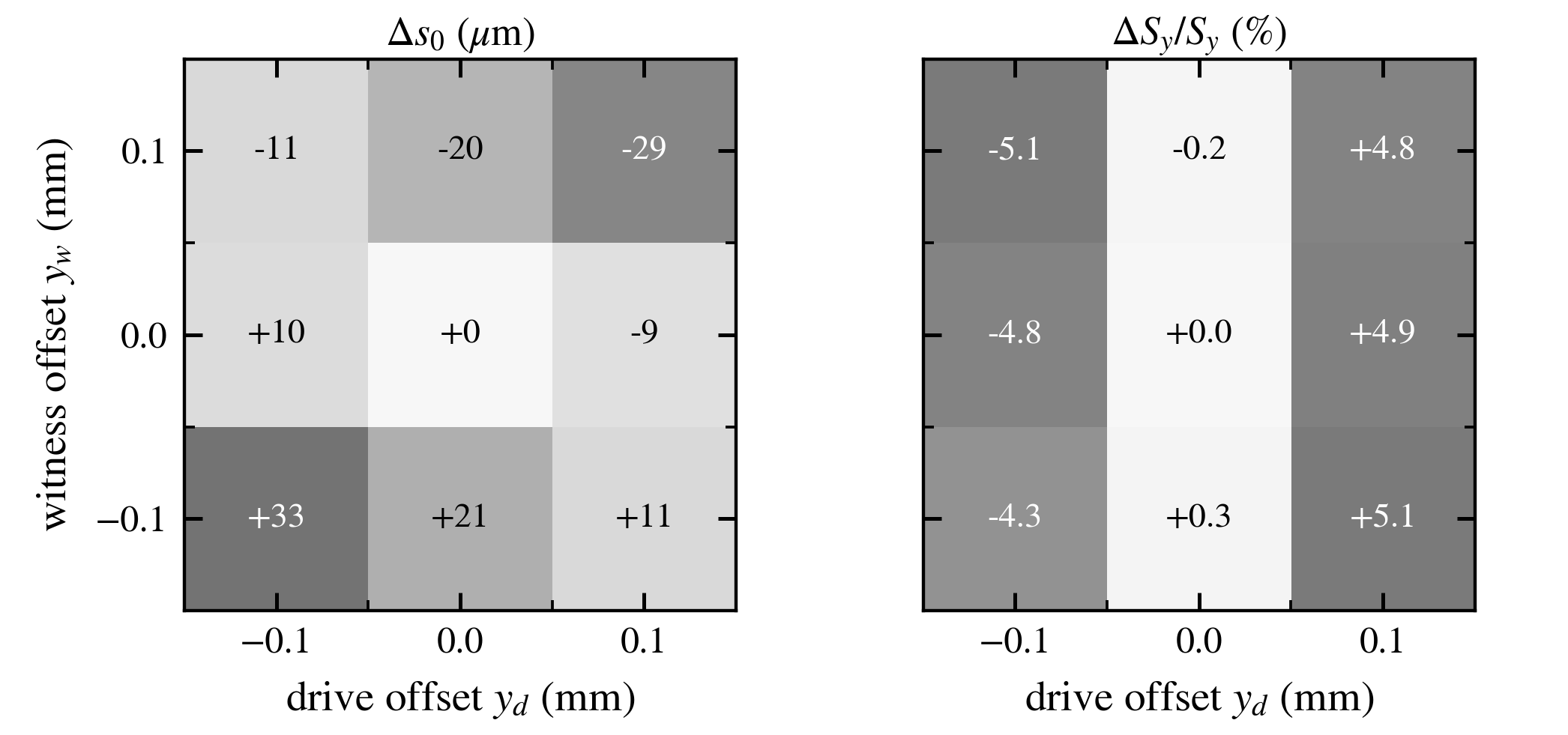}
  \caption{
  Sensitivity of the transverse-wake zero crossing and local temporal slope to
  independent vertical orbit offsets.  The horizontal axis is the drive-bunch
  offset and the vertical axis is the witness-bunch offset, both given as
  offsets from the nominal trajectory.  The left panel shows the shift of the
  zero crossing \(s_0\), while the right panel shows the relative change of the
  local slope.
  }
  \label{fig:orbit-sensitivity}
\end{figure}

The orbit sensitivity was evaluated by varying the drive and witness
trajectories in the vertical plane using the 5-cell model.  For
each case, the zero crossing \(s_0\) closest to the nominal witness position was
found by linear interpolation and the local slope was extracted from a local
linear fit to \(W_y(s)\) around that crossing.
Figure~\ref{fig:orbit-sensitivity} summarizes the resulting shift of the zero
crossing and the corresponding change in streaking strength.

For the tested \(\pm\SI{100}{\micro m}\) vertical offsets, the local temporal
slope changes by less than about \(6\%\).  The zero-crossing position is more
sensitive, with shifts up to about \(\SI{30}{\micro m}\).  With the sign
convention used in the simulations, negative vertical offset moves the
trajectory toward the pillbox axis.  The tested opposite horizontal split
\((x_d,x_w)=(-\SI{100}{\micro m},+\SI{100}{\micro m})\) changes the vertical
wake negligibly, with \(\Delta s_0=-\SI{1.0}{\micro m}\) and
\(\Delta S_y/S_y=-0.02\%\).  

\section{\label{sec:tuning}Tuning}

Uniform scaling of the copper geometry was used to estimate the sensitivity of
the wake phase to structure temperature.  The model assumes homogeneous thermal
expansion and does not include thermal gradients or cooling-channel geometry.
For copper, the scale factor is
\begin{equation}
  \lambda_T
  =
  1+\alpha_{\rm Cu}\Delta T ,
  \qquad
  \alpha_{\rm Cu}=\SI{16.5e-6}{K^{-1}} .
  \label{eq:thermal-scale-factor}
\end{equation}
Thus \(\Delta T=\pm\SI{10}{\celsius}\) corresponds to
\(\lambda_T=0.999835\) and \(\lambda_T=1.000165\).  The scaled geometries were
compared with a reference case at \(\lambda_T=1\).

\begin{figure}[t]
  \centering
  \includegraphics[width=0.65\linewidth]{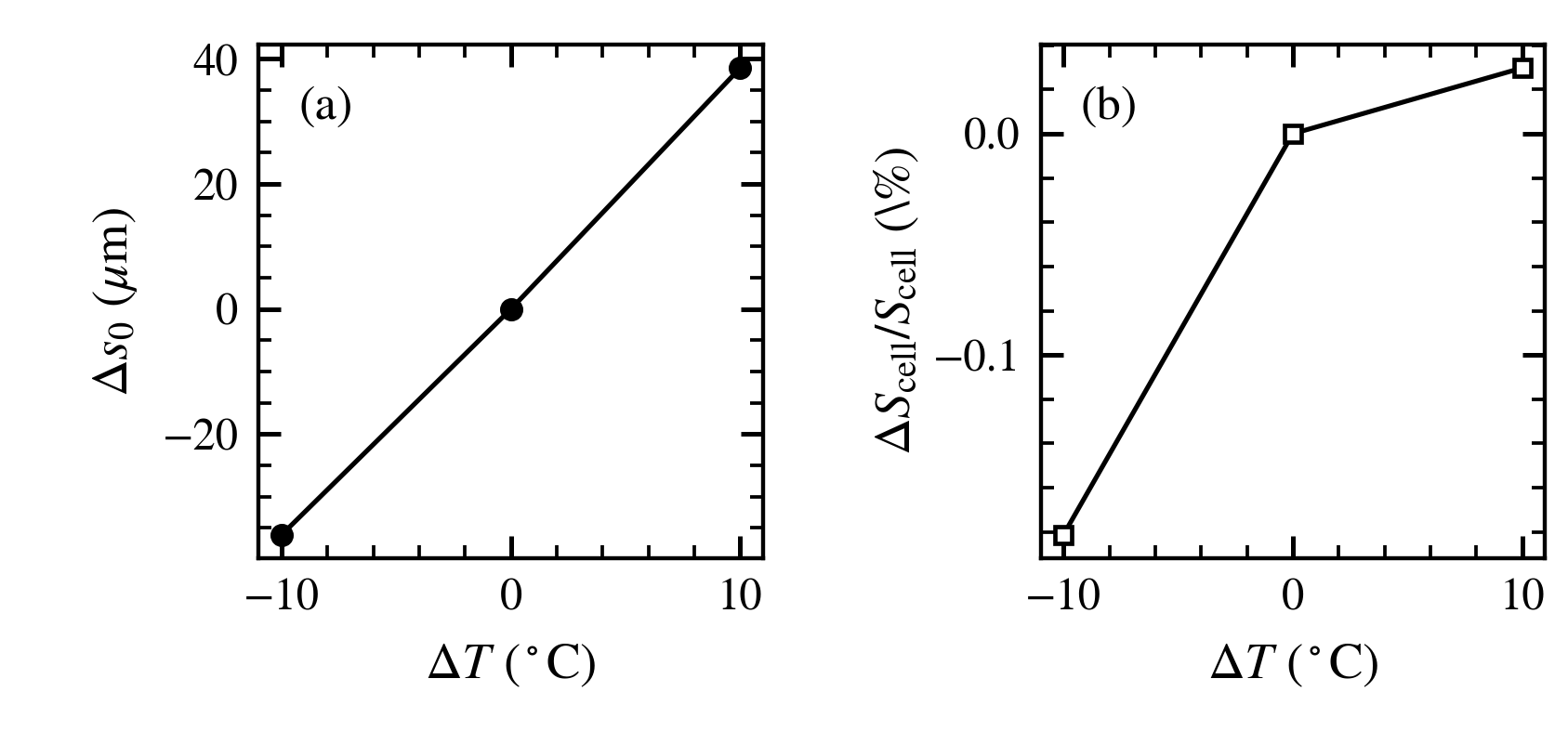}
  \caption{
  Uniform thermal-scaling sensitivity of the wake operating point.  Panel (a)
  shows the shift of the zero crossing relative to the reference geometry
  \(\lambda_T=1\).  Panel (b) shows the relative change of the per-cell temporal
  slope \(S_{\rm cell}\).  The scaled cases assume
  \(\lambda_T=1+\alpha_{\rm Cu}\Delta T\), with
  \(\alpha_{\rm Cu}=\SI{16.5e-6}{K^{-1}}\).
  }
  \label{fig:water-tuning}
\end{figure}

The result is shown in Fig.~\ref{fig:water-tuning}.  The zero crossing shifts
approximately linearly with temperature:
\[
  \Delta s_0(-\SI{10}{\celsius})=-\SI{36.1}{\micro m},
  \qquad
  \Delta s_0(+\SI{10}{\celsius})=+\SI{38.6}{\micro m}.
\]
 In contrast, the local
streaking strength remains essentially unchanged: the relative changes of
\(S_{\rm cell}\) are \(-0.18\%\) and \(+0.03\%\).  Therefore, uniform thermal
scaling mainly shifts the witness phase while preserving the local wake slope
over the tested \(\pm\SI{10}{\celsius}\) range.

A controlled collinear orbit offset provides a second phase-tuning mechanism. We
extend the study to collinear drive--witness offsets over the range
\(-\SI{300}{\micro m}\le y \le \SI{300}{\micro m}\).

\begin{figure}[!htb]
  \centering
  \includegraphics[width=0.55\linewidth]{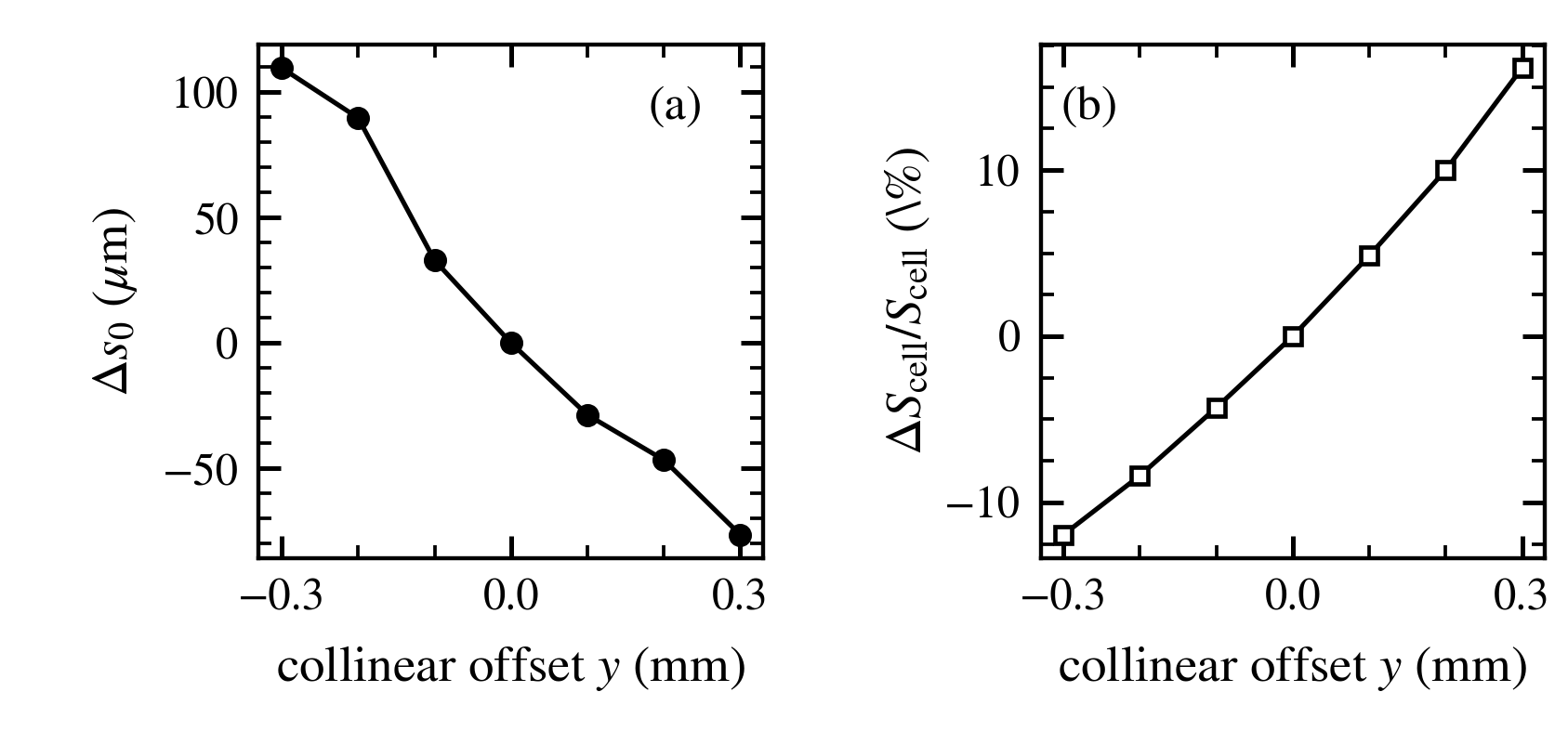}
  \caption{
  Collinear orbit-offset dependence of the transverse wake.  The drive
  and witness trajectories are displaced together with respect to the nominal
  beam-tube axis.  Panel (a) shows the zero-crossing shift
  \(\Delta s_0=s_0(y)-s_0(0)\).  Panel (b) shows the corresponding relative
  change of the per-cell temporal slope \(S_{\rm cell}\).
  }
  \label{fig:collinear-offset-tuning}
\end{figure}

Figure~\ref{fig:collinear-offset-tuning} shows the effect of a common
collinear displacement of the drive and witness trajectories.  The nominal
case is defined by \(y=0\), for which
\[
 s_0\simeq\SI{230.6}{mm},
  \qquad
  S_{\rm cell}=\SI{1.186}{mV/(pC\,fs\,cell)} .
\]
Over the tested range
\(-\SI{0.3}{mm}\le y\le\SI{0.3}{mm}\), the zero crossing shifts
monotonically from \(+\SI{110}{\micro m}\) to
\(-\SI{77}{\micro m}\) relative to the nominal case.  With the present sign
convention, negative \(y\) moves the trajectory toward the pillbox axis.  The
same scan changes the per-cell temporal slope by
\(-\SI{12}{\percent}\) to \(+\SI{16}{\percent}\) relative to nominal,
corresponding to absolute values from
\(\SI{1.049}{mV/(pC\,fs\,cell)}\) to
\(\SI{1.384}{mV/(pC\,fs\,cell)}\).

\subsection{Movable parts}
 Unlike conventional high-power rf resonators, a passive resonant deflector does not require rf contact between two halves of the structure. In addition to simplified production and cleaning, this can be interpreted as an additional tuning mechanism. The acceptance of the split gap is explained by the surface-current topology of
the dominant TM-like modes.  On the cylindrical wall of a pillbox-like TM mode,
the azimuthal magnetic field produces an axial surface current,
\[
  \mathbf K_s=\hat{\mathbf r}\times H_\phi\hat{\boldsymbol\phi}
  =H_\phi\hat{\mathbf z}.
\]
For a longitudinal structure split, this current flows along the seam.  The gap therefore acts as a controlled perturbation of the eigenmode spectrum.  

\begin{figure}[!htb]
  \centering
  \includegraphics[width=0.35\linewidth]{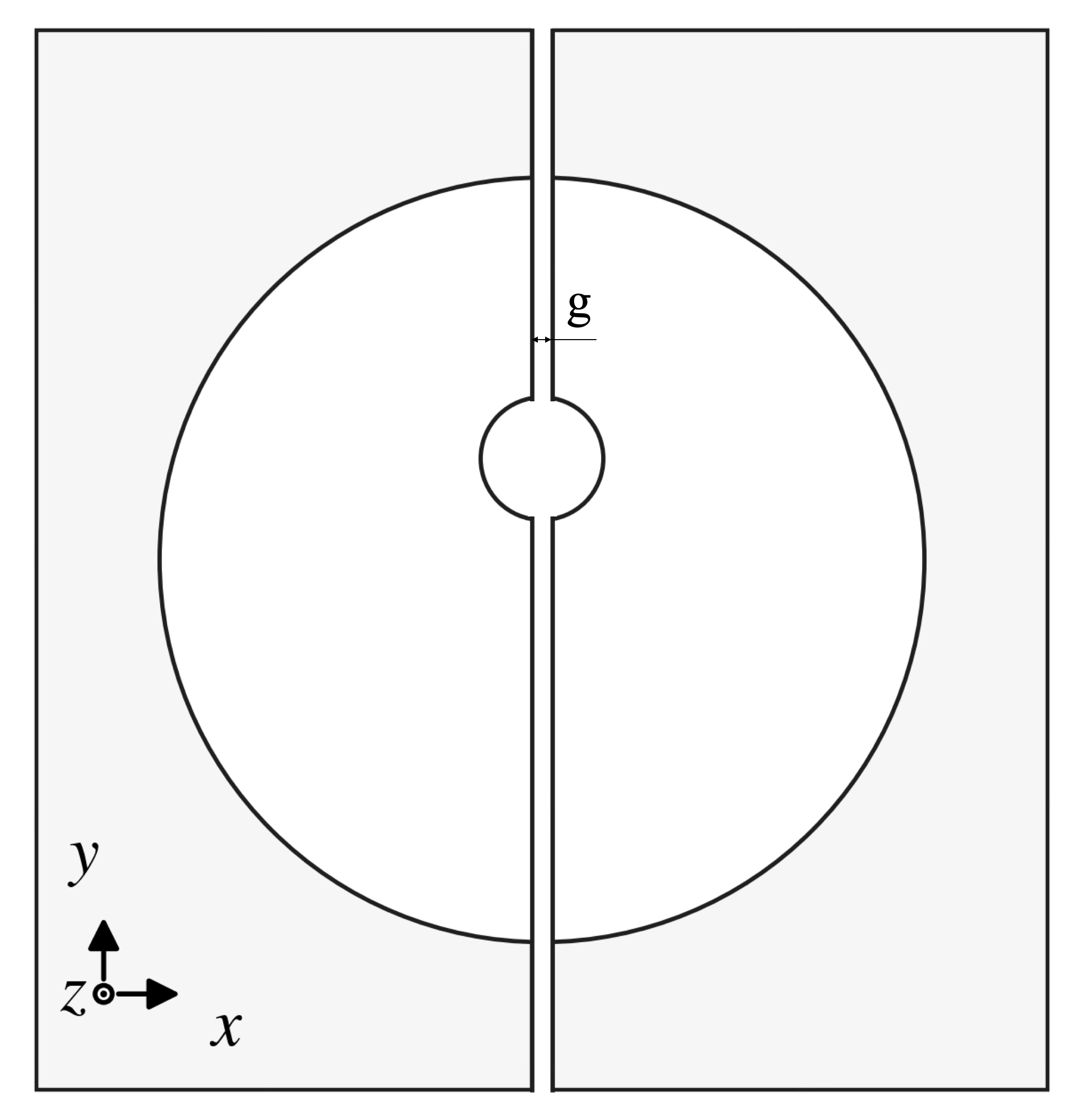}
  \caption{
  Schematic \(x\)--\(y\) cross section of the split-block off-axis pillbox
  structure.  The beam-pipe aperture is displaced in \(y\) with respect to the
  pillbox axis and the beam direction \(z\).
  }
  \label{fig:gap-schematic}
\end{figure}

\begin{figure}[!htb]
  \centering
  \includegraphics[width=0.8\linewidth]{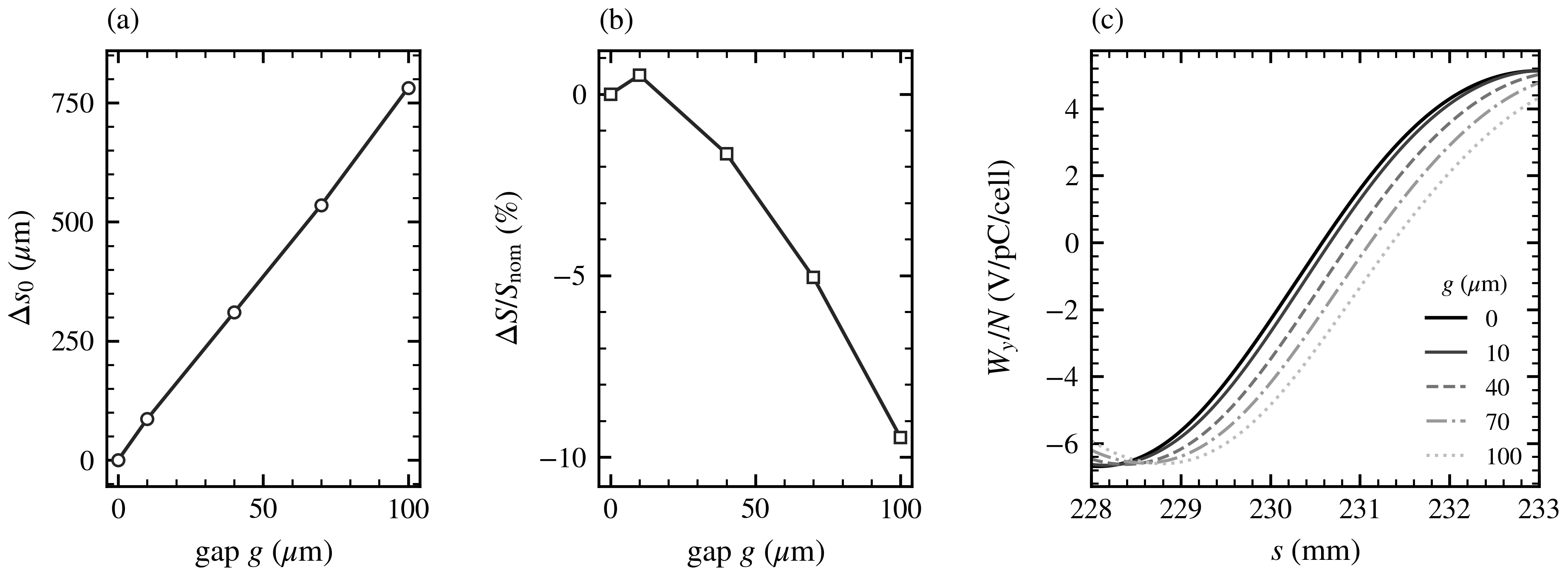}
  \caption{
  Mechanical gap sensitivity of the split-block off-axis pillbox structure.
  (a) Schematic \(x\)-\(y\) cross section of the metallic enclosure of the deflector,
  where \(z\) is the beam direction.  (b) Shift of the transverse-wake zero
  crossing relative to the zero-gap nominal structure.  (c) Relative change of
  the temporal streaking slope, \(\Delta S/S_{\rm nom}\), evaluated at the
  corresponding zero crossing for each gap.  (d) Wake
  potentials over the same absolute witness-delay interval. 
  }
  \label{fig:gap-study}
\end{figure}

The split-block geometry is shown in Fig.~\ref{fig:gap-schematic}.  A
 gap of width \(g\) is introduced between the two copper halves in
the \(x\)-direction.  Figure~\ref{fig:gap-study} shows the resulting change of
the wake operating point.  For
\(\SI{10}{\micro\meter}\le g\le\SI{100}{\micro\meter}\), the zero crossing
\(s_0(g)\) shifts monotonically to larger drive--witness separation.  Relative
to the closed geometry, the shift reaches approximately
\(\SI{0.78}{mm}\) at \(g=\SI{100}{\micro\meter}\).

The local temporal slope was evaluated at the zero crossing corresponding to
each gap.  Over the same range, \(S_{\rm cell}(g)\) changes moderately: it is
nearly unchanged at \(g=\SI{10}{\micro\meter}\) and is reduced by about
\(\SI{9.5}{\percent}\) at \(g=\SI{100}{\micro\meter}\).  The wake traces in
Fig.~\ref{fig:gap-study}(d) show that this perturbation mainly changes the
phase of the wake packet.  The zero crossing is displaced in
\(s\), while the local linear region used for streaking is retained over the
tested gap range. An additional scan up to \(g=\SI{500}{\micro m}\) gives the same sign of the
phase shift, but no longer belongs to the small-perturbation regime.  The zero
crossing then shifts by millimetres and the local slope decreases
accordingly. 

\section{European XFEL low-energy use case example}
\label{sec:b1-use-case}

The optimized wake was applied to a longitudinal phase-space diagnostic example
in the European XFEL~\cite{Decking2020} after the second bunch compressor in the
three-bunch-compressor layout.  The beam optics determine the screen
calibration and the resolution through Eq.~\eqref{eq:temporal-resolution}.  Here, we use a witness energy \(E_w=\SI{700}{MeV}\), drive charge
\(q_d=\SI{250}{pC}\) and horizontal transfer element
\(R_{12}=-\SI{38.73}{m}\) from the structure to the observation screen.  For this particle tracking example the same structure is rotated so that the selected streaking plane is horizontal, with \(u_s=x\), \(R_s=R_{12}\) and
\(\sigma_{u_s,0}=\sigma_{x,0}\).  Tracking without streaking gives an
unstreaked horizontal rms beam size \(\sigma_{x,0}=\SI{122.2}{\micro m}\).  

The nominal 5-cell wake gives a per-cell streaking strength
\[
  S_{\rm cell}
  =
  \SI{1.186}{mV/(pC\,fs\,cell)} .
\]
Using the period \(P=\SI{4.42}{mm}\), a structure of length
\(L=\SI{1}{m}\) contains  \(N=L/P=226\)~cells.  The
corresponding temporal slope is
\[
  N S_{\rm cell}
  \simeq
  \SI{0.27}{V/(pC\,fs)} .
\]
Inserting this value into Eq.~\eqref{eq:temporal-resolution} gives
\begin{equation}
  \sigma_{t,\rm res}
  \simeq
  \SI{33}{fs}
  \label{eq:b1-use-case-resolution}
\end{equation}
for the B1 optics and the tracked unstreaked beam size.  

\begin{figure}[t]
  \centering
  \includegraphics[width=\linewidth]{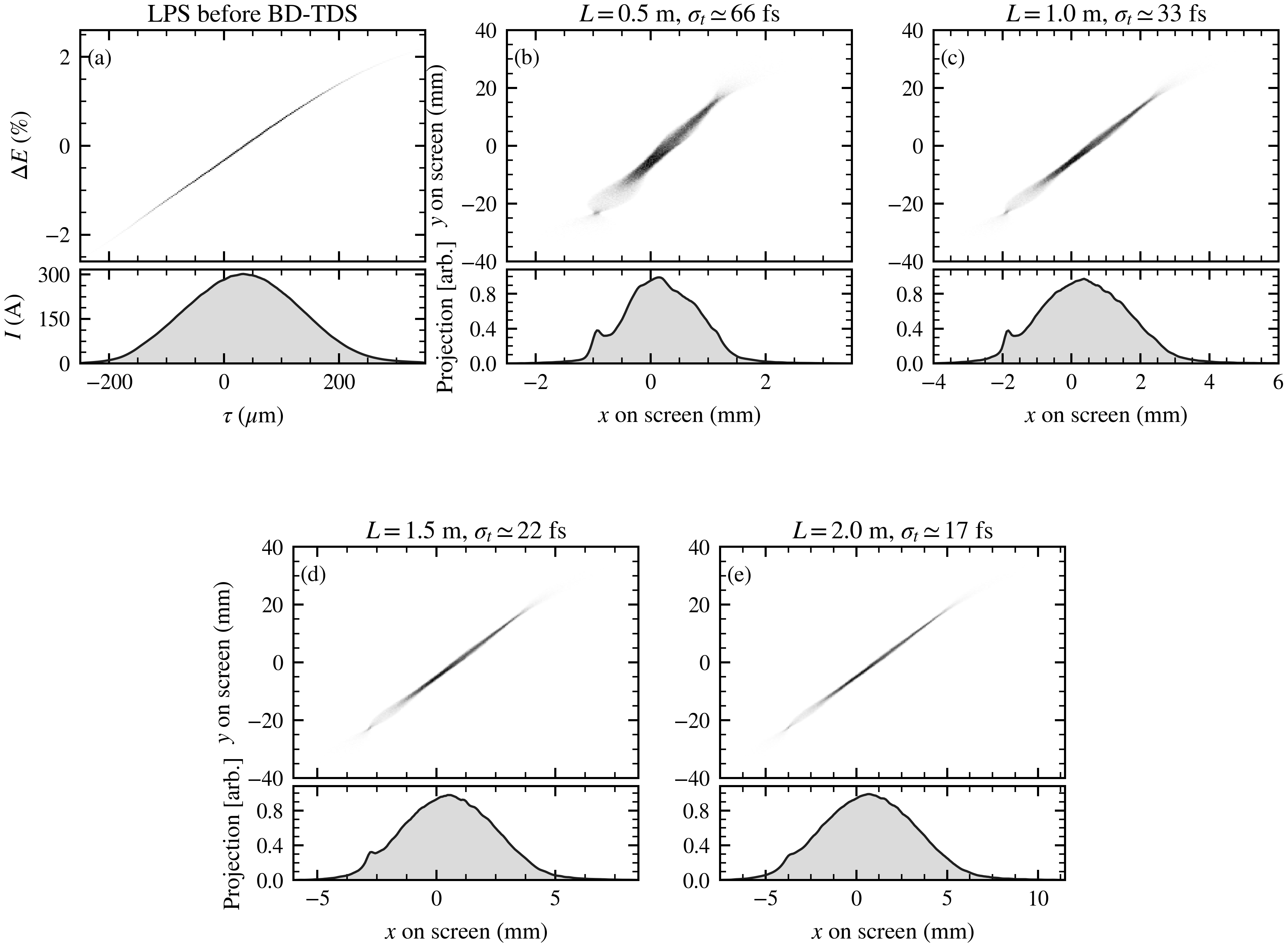}
\caption{
Beam-dynamics example for a \(\SI{3}{mm}\)-aperture beam-driven passive
deflector in the B1 section.  Panel (a) shows the incoming longitudinal phase
space and current profile before the structure.  Panels (b)--(e) show the
simulated screen distributions and horizontal projections for structure lengths
of \(\SI{0.5}{m}\), \(\SI{1.0}{m}\), \(\SI{1.5}{m}\) and \(\SI{2}{m}\) respectively.   The particle tracking includes the drive-bunch-induced transverse and longitudinal wakes, as well as the longitudinal self-wake of the witness bunch.
}
  \label{fig:b1-use-case-tracking}
\end{figure}

Tracking simulations were then performed for \(\SI{0.5}{m}\),
\(\SI{1.0}{m}\)), \(\SI{1.5}{m}\)) and \(\SI{2}{m}\) long deflecting structures (see Fig.~\ref{fig:b1-use-case-tracking}).  The wakefields
were applied as external kicks and energy changes: the long-range transverse
and longitudinal wakefields were driven by the \(\SI{250}{pC}\) drive bunch and
the longitudinal self-wake of the witness bunch was included separately.  For a
\(\SI{1}{m}\) structure, the longitudinal long-range wake from the drive bunch
and the longitudinal short-range self-wake of the witness bunch each produce an
energy modulation of order \(\SI{2}{MeV}\), corresponding to a few \(10^{-3}\)
of the \(\SI{700}{MeV}\) witness energy.  The transverse streak is produced by
the drive-bunch-induced transverse wake.  The longitudinal wakes change the
witness energy distribution and are relevant for the final screen image when
the observation optics is dispersive.  Their deterministic contribution can be
included, in principle, in the image analysis or corrected during reconstruction.

The simulated screen separation increases with structure length, as expected
from the approximately constant per-cell slope.  For the present
\(\SI{3}{mm}\)-class aperture, the \(\SI{1}{m}\) case gives an estimated
temporal resolution of about \(\SI{33}{fs}\) for the stated optics and charge.
For otherwise similar geometries, reducing the aperture is expected to increase
the wake slope but reduce alignment margin. It is worth noting that the estimates above assume a drive charge of \(\SI{250}{pC}\).  The European XFEL linac can transport higher drive-bunch charges, for example \(\SI{500}{pC}\), which would increase the streaking strength and therefore improve the temporal resolution linearly with charge.

\FloatBarrier

\section{Conclusion}
\label{sec:conclusion}

We formulated a modal design framework for a beam-driven resonant passive
transverse deflector and evaluated it with wake simulations.  The useful
operating point is a long-range transverse-wake zero crossing and the
streaking strength is characterized by the corresponding local temporal slope
\(S_s=c\,dW_s/ds\). The relevant modal weight
is the product of drive longitudinal voltage and witness transverse Lorentz
voltage, \((V_z^d)^*V_s^w/U_m\).  This explains why off-axis TM-like parent families can provide a
streaking wake: the drive bunch excites them through \(E_z\), while the witness
samples the transverse Lorentz force associated with the same modal fields, provided mostly by the magnetic component.

For the optimized finite structure, the 5-cell wake gives
\(s_0\simeq\SI{230.6}{mm}\) and
\(S_{\rm cell}=\SI{1.186}{mV/(pC\,fs\,cell)}\).  The 5-cell and 56-cell
comparison shows that the per-cell temporal slope is nearly unchanged over the
simulated range, indicating coherent accumulation of the useful derivative.
A wake-only harmonic decomposition shows that the response is composed of
multiple   up to 75 GHz.

The sensitivity analysis gives initial tolerance estimates for the optimized
geometry.  In the tested ensemble, the sample-mean slope reduction for
\(\SI{10}{\micro m}\) rms random cell perturbations is
\(\SI{5.5}{\percent}\).  Independent drive--witness orbit offsets mainly shift
the wake phase, while collinear
orbit offsets change both phase and modal overlap.  Uniform copper scaling over
\(\pm\SI{10}{\celsius}\) shifts the zero crossing by about
\(\pm\SI{40}{\micro m}\) with negligible slope change in the simulated model. A longitudinal split-block gap provides a stronger phase perturbation: for
\(\SI{10}{\micro m}\le g\le\SI{100}{\micro m}\), the zero crossing moves
monotonically by up to \(\SI{0.78}{mm}\), while the local slope remains within
about \(\SI{10}{\percent}\) of the closed-geometry value.
For the European XFEL low-energy example, a \(\SI{1}{m}\) structure, a
\(\SI{250}{pC}\) drive bunch and a \(\SI{700}{MeV}\) witness beam give an
estimated temporal resolution of about \(\SI{33}{fs}\).  

One of the central practical conclusions is that the high-frequency multi-mode wake
composition does not make the operating point fragile.  In principle, small
geometric or orbit perturbations could change the modal phases and suppress the
zero-crossing derivative.  The simulations show instead that the tested
perturbations mainly displace the wake phase or change the modal overlap
moderately, while preserving a usable local slope.  The temperature, orbit and
split-gap correlations provide tuning flexibility for setting the
working point together with beam-based calibration.  The split-block result is
especially relevant for implementation as the structure does not require an RF-tight electrical contact between the two copper halves. Water cooling can be used for temperature and phase stabilization.

\bibliography{apssamp}

\end{document}